# Forecasting the impact of COVID-19 on health workforce supply in EU countries

Peter Klimek[1,2], Katharina Ledebur[1,2], Michael Gyimesi[3], Herwig Ostermann[3,4], Stefan Thurner[1,2,5,6,*]

[1]Section for Science of Complex Systems, CeMSIIS, Medical University of Vienna, Spitalgasse 23, A-1090 Vienna, Austria;

[2]Complexity Science Hub Vienna, Josefstädter Straße 39, A-1080 Vienna, Austria;

[3]Austrian National Public Health Institute, Stubenring 6, A-1010 Vienna, Austria;

[4]Department for Public Health, Health Services Research and HTA, UMIT – Private University for Health Sciences, Medical Informatics and Technology, EWZ1, A-6060, Hall in Tirol, Austria;

[5]IIASA, Schlossplatz 1, A-2361 Laxenburg, Austria;

[6]Santa Fe Institute, 1399 Hyde Park Road, NM 85701, USA

[*]stefan.thurner@muv.ac.at

## Abstract

Many countries faced challenges in their health workforce supply like impending retirement waves, negative population growth, or a suboptimal distribution of resources across medical sectors even before the SARS-CoV-2 pandemic struck. We propose a novel population-dynamical and stock-flow-consistent approach to health workforce supply forecasting that is complex enough to address dynamically changing behaviors while requiring only publicly available timeseries data for complete calibration. We demonstrate the usefulness of this model by applying it to 21 European countries to forecast the supply of generalist and specialist physicians until 2040, as well as how Covid-related mortality and increased healthcare utilization might impact this supply. Compared to staffing levels required to keep the physician density constant at 2019 levels, we find that in many countries there is indeed a trend toward decreasing density for generalist physicians at the expense of increasing densities for specialists. The trends for specialists are exacerbated in many Southern and Eastern European countries by expectations of negative population growth. We expect a limited impact of Covid on these trends even under conservative modelling assumptions. Finally, we generalize the approach to a multi-professional, multi-regional and multi-sectoral model for Austria where we find an additional suboptimal distribution in the supply of contracted versus non-contracted (private) physicians. It is therefore of the utmost importance to devise tools for decision makers to influence the allocation and supply of physicians across fields and sectors to combat these imbalances.

## Significance Statement

We develop a parsimonious modelling framework to project over the next twenty years whether 21 European countries have to expect increasing or decreasing supply of generalist or specialist physicians, as well as how post-Covid disruptions might impact on these densities. We find a trend toward decreasing density for generalist physicians at the expense of increasing densities for specialists, driven by changes in the age structure of the physicians, the general population, as well as the expected numbers of graduating and migrating physicians. Compared to these factors, we find a moderate impact of Covid-related disruptions that are typically smaller than the uncertainties associated with the projections of the physician densities.

# Introduction

Health workforce planning (HWFP) has the objective to achieve a balance between supply and demand of health workers [1]. There are several health sector-specific challenges that need to be addressed in HWFP [2]. First, medical professionals require substantial investments of time and costs to be trained [3]. Second, demographic and socio-economic changes might impact on the demand for healthcare services in a way that is hard to anticipate [4]. Third, retirement patterns of physicians are highly dependent on their region, sex, and other personal factors that can also change over time [5]. Fourth, to quantitatively assess whether there is a gap between supply and demand in the future (and whether this gap is increasing or narrowing), data with enough quality must be available to adequately assess the status quo in the numbers of health professional [2]. Fifth and finally, many healthcare systems face tight fiscal constraints which exacerbates the need for more accurate HWFP [6]. Overall, countries have limited policy options to steer their health workforce supply to address these challenges. In practice, they either try to influence the number of new graduates (e.g., introducing a *numerus clausus*) [7] or (ii) adjust the number of contracts offered or auctioned to physicians [8].

In the years before the Great Recession there was a widespread consensus amongst OECD countries that demand for physicians was increasing more rapidly than their supply [9]. This led to policy recommendations to increase the health workforce in several countries. However, together with the slowdown of economic activity and reduced willingness for health expenditures in the population, concerns soon shifted towards a potential oversupply of physicians in certain areas [4]. Several countries (for instance, from Southern and Eastern Europe) face negative population growth or have an age structure of their health workforce that is strongly skewed toward older ages [10]. With these impending waves of retirements, there have also been concerns for physician shortages [11].

Uncertainty in the field of health workforce planning is further increased due to potential long-term impacts of the SARS-CoV-2 pandemic. With the virus transitioning into endemicity, there is concern that healthcare demand might increase due to long-lasting symptoms of infections [12]. Furthermore, health personnel were at an increased risk of becoming infected which means that risks for Covid-related mortality or work disability are particularly high in this occupational group [13], potentially leading to work disability [14]. In this work we seek to quantify the extent to which these stressors might impact the balance of future supply and demand of health workforce in European countries.

Several methodological approaches have been proposed to quantitatively forecast changes in supply and demand in the health workforce [1, 4, 15, 16, 17]. Two frequently encountered approaches are linear extrapolation and stock flow consistent modelling. The main idea in the linear extrapolation approach is to consider historical timeseries of the numbers of healthcare professionals (graduates, physicians, etc.), measure their linear trend (potentially adjusted for confounding variables) and to extrapolate this trend into the future [15, 18, 19, 20]. Advantages of this approach include minimal data requirements (historical timeseries are often enough) and low complexity of the underlying model (e.g., linear regression). However, without adding additional assumptions this approach fails to capture changing behavior amongst physicians or in the population and – even worse – might in some cases lead to nonsensical models where graduates of physicians (dis)appear out of or into nowhere. Linear extrapolation methods require further modifications or ad-hoc assumptions to account for disruptive events like the SARS-CoV-2 pandemic, which cannot be modelled by simply extrapolating the future from the past. A more sophisticated approach that addresses some of these shortcomings is the use of stock flow consistent models [21, 22]. The key idea here is to represent all relevant types of health professionals as separate entities in the model (physicians in different fields, graduates, migrants, etc.) and to explicitly include all in and out flows of all entities, as well as all flows between them. These models are stock-flow-consistent in the sense that no doctors appear out of thin air. Their main shortcoming is their high dimensionality. In principle, it is necessary to specify every flow rate in the model (which can easily go into dozens of free parameters, let alone the specification of how these parameters might change). In practice, stock flow consistent models require extensive data to be meaningfully calibrated, often on the level of individuals [23]. Both approaches can be conducted at different levels of sophistication, ranging from single to multiple professions, geographic regions, and health sectors [7, 16].

In this work we introduce a new approach to HWFP models that aims to combine the strengths of the linear extrapolation and stock flow consistent modelling approaches while circumventing some of their shortcomings. Our approach is inspired by population dynamical models that are routinely employed in scientific fields such as evolutionary game theory [24], ecology [25], or the theory of complex systems [26]. Like the stock-flow-consistent approach, different types of health professionals are represented by separate entities in the model. Instead of specifying all flow rates between the model entities, we are only interested in the net rates of change between them. This brings the key advantage that these net rates can be directly estimated from timeseries data—resulting in a model that has a minimal number of free parameters. Our approach therefore has similar data requirements compared to linear extrapolation models while keeping stock flow consistency intact (i.e., no uncontrollable sources or sinks in the model). Additionally, we ensure that all flow rates that are typically used for policy interventions (e.g., number of graduates, positions offered to physicians) are explicitly represented in the model such that it is possible to quantitatively assess the impact of intervention that target these rates.

Note that in this work we explicitly focus on health workforce supply. We do not attempt to forecast demand per se but rather use demographic projections of population growth to estimate the future supply needed to keep the physician density constant at 2019 levels. Additionally, we will consider model scenarios for how the pandemic might influence this demand under conservative assumptions.

To introduce and study the model we will proceed as follows. First, we introduce a minimal version of the model that can be fully calibrated with publicly available data for 21 European countries. For these countries we forecast supply of general practitioners (GPs) and medical specialists. We compare these projections with the supply required for constant physician density and explore how this gap is projected to develop until 2040 across different European regions. For the case of Austria, we also show how the minimal model can be extended to a multi-sectoral and multi-professional model that takes country-specific properties of the healthcare system into account, without having to introduce free ad hoc parameters. We then formulate conservative scenarios for the long-term impact of the SARS-CoV-2 pandemic on the health workforce and quantify how many additional physicians might be required because of this shock.

The minimal model requires the following data as input for a given country: timeseries of the number of graduates, numbers of immigrating physicians, the number of GPs, and the number of medical specialists, stratified by age and sex. This data is available with sufficient quality for 21 European countries in the EUROSTAT database for the timespan 2000-2019, see Methods.

The structure of the minimal model is shown in Figure 1. We consider graduates, migrants, GPs, and other medical specialists (not including pediatricians and dentists). GPs and specialists are characterized by their age and sex. Assume we observe a certain stock of GPs with a given sex and age in year $t$. If none of these GPs enter or leave the system within the next, say, 10 years, we would expect to have an equal stock of GPs that are ten years older after ten years. This means that an effective or net rate of change for GPs of a given age and sex can be computed by comparing these two numbers, see Methods. A plethora of processes contributes to this net rate of change: physicians retire, move, migrate, change field or profession, and so forth. However, for the purpose of HWFP supply modelling, it is primarily of interest to us at which rate they *effectively* enter or leave the system.

The model is implemented by repeatedly iterating through three update steps. In the first step, see Figure 1A, B, physicians enter the health care system either via graduation or immigration with a certain probability, denoted by $p_{\text{enter}}$, see Methods. The number of future graduates and migrants is estimated as the average of the three most recent numbers. New physicians (graduates or migrants) enter the system with probability $p_{\text{GP}}$ as a GP and with probability $1 - p_{\text{GP}}$ as a specialist. The parameters $p_{\text{enter}}$ and $p_{\text{GP}}$ are in general not known but can be estimated from data, see Methods. In the second step of the model, physicians age, see Figure 1C, D, where we show the age pyramid for (C) female and (D) male GPs in Austria in 2016. In the third step, physicians exit with age- and sex-specific rates that are estimated from data, see Figure 1E, F and Methods. The execution of steps 1-3 constitutes one iteration of the model and corresponds to one year in real time.

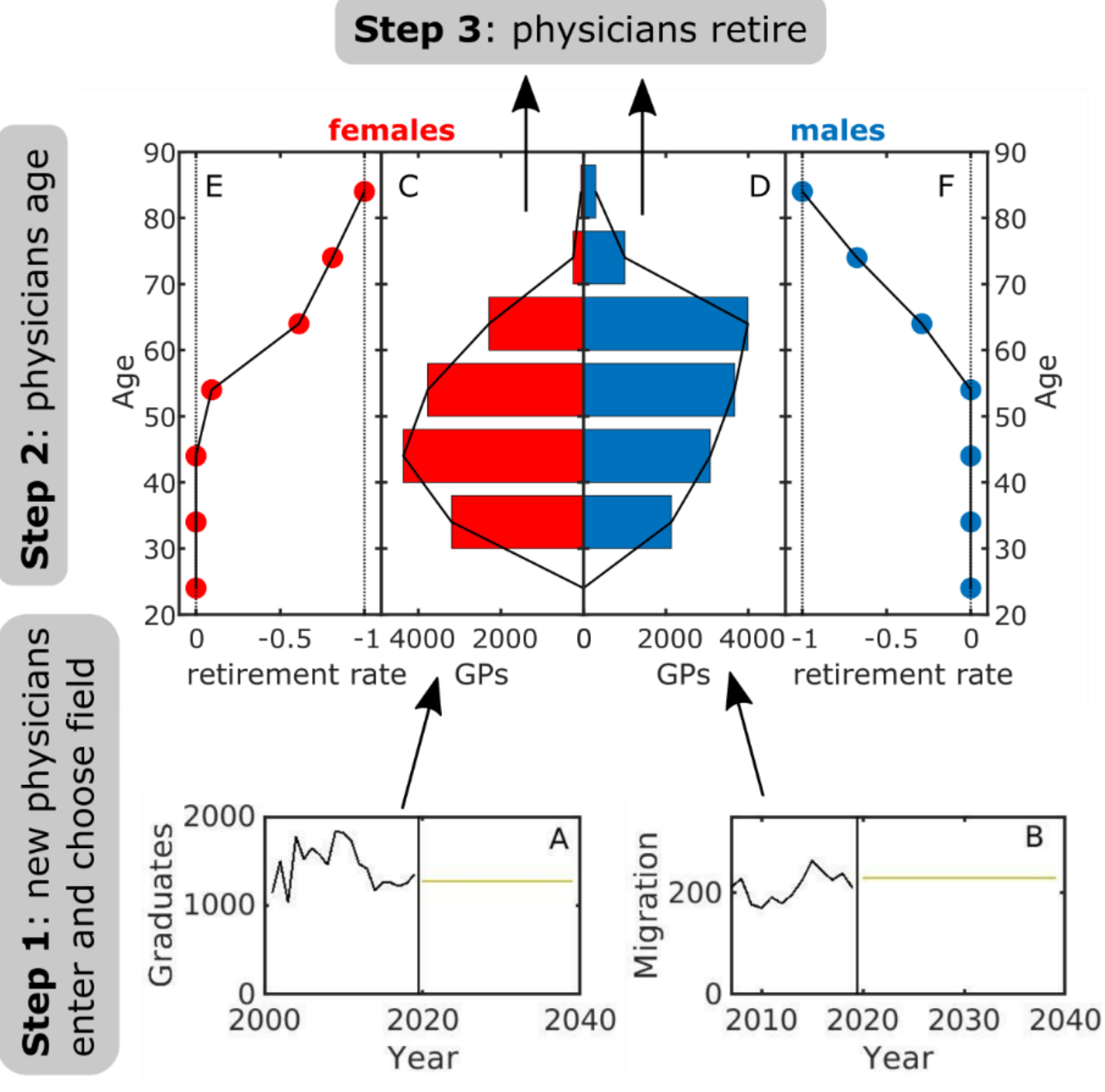

*Figure 1: Structure of the minimal model. The model dynamics consists of three steps. In the first step, (A) graduates or (B) migrants enter the healthcare system and choose a specific field, here they enter the stock of GPs. Next to their field, physicians are also characterized by their sex and age distribution, see (C) for females and (D) females. In the second step of the model dynamics physicians age by climbing up in the age pyramid. Finally, in step 3, physicians retire with age- and sex-specific rates, see (E,F). The data shown in this example was taken from Austria in 2016.*

The implementation of the model occurs in three phases: calibration, validation, and forecast. At calibration, the initial number of physicians and their demographic structure are taken from the data for 2000. In the calibration phase, the update steps 1-3 are applied for the years 2001-2019. The parameters $p_{\text{enter}}$ and $p_{\text{GP}}$ are chosen such that the model timeseries for the numbers of GPs and specialists best match the observed timeseries (goodness-of-fit is evaluated by the sum of chi-squared distances between data and model physician numbers, see Methods). Using these values, the model is iterated until 2040 in the forecast phase. See Methods for a full protocol of the model.

The forecasted supply is compared to the number of physicians that would be required to keep the physician density constant and 2019 levels. These densities are estimated from the baseline population growth scenarios provided by EUROSTAT, together with an envelope for alternative growth scenarios for high or low migration or fertility, see Methods.

The primary outcome of the model is the yearly gap between the current and future physician density computed over 2020-2040, the density gap $DG$. To make $DG$ comparable across countries, we measure it relative to the average number of graduates in a country, see Methods. The value of $DG$ has the precise meaning of percentage of graduates that need to be added ($DG > 0$) or removed ($DG < 0$) per year in the model for constant physician density.

The implementation of the Covid shock in the model is based on the assumption that the demand for physicians in the population increases due to long-term sequelae of severe cases (patients who were admitted to ICU) and due to patients with long-lasting Covid symptoms who were not admitted to the hospital. On the other hand, the supply of physicians decreases due to work inability of physicians who experienced a severe Covid course. We study a conservative, worst-case scenario that assumes that the prevalence of severe cases and long-lasting symptoms in the upcoming years will be equal to the annual prevalence observed in the first two years of the pandemic.

# Results

**Fitting parameters.** We show the output of the minimal model for four different countries in Figure 2, namely for (A-C) Austria, (D-F), Croatia, (G-I) Netherlands, and (J-L) Great Britain. First, we show results for the estimated parameters $p_{\text{enter}}$ and $p_{\text{GP}}$ in the first row, panels (A, D, G, J). Three observations can be made here, that also apply to most other countries, see Figures S1-S3. First, we see that the model has a clearly defined global optimum in terms of goodness-of-fit (the yellow areas indicating parameter values of low chi-squared distance between data and model). Second, we observe as a general trend that the probabilities $p_{\text{enter}}$ to enter the system after graduation are close to one, but not necessarily one. Third, we find rates of entering as a GP that vary substantially across countries, between 10% for Croatia to more than 40% in the Netherlands.

**Forecasts.** In the second and third row in Figure 2 we show results for the data and model timeseries of the numbers of GPs (second row) and specialists (third row). These plots can be read as follows. At initialization, the model is calibrated to data form the first year, in most cases 2000, indicated by the red circles where data and model physician numbers are identical. The black timeseries span over the validation phase until 2019, where we compare the data (dotted lines) with model timeseries (solid line), error bars give the root-mean-squared errors (RMSE) between model and data. From 2020 onwards (to the right of the black vertical line) the forecasts are shown in yellow. The solid line gives the model forecast for the numbers of physicians; error bars show projected RMSE by means of Gaussian error propagation. The yellow dotted line gives the number of physicians that would be required for a density at constant 2019 levels given the baseline population growth scenarios; the yellow shaded areas envelop all alternative population growth scenarios.

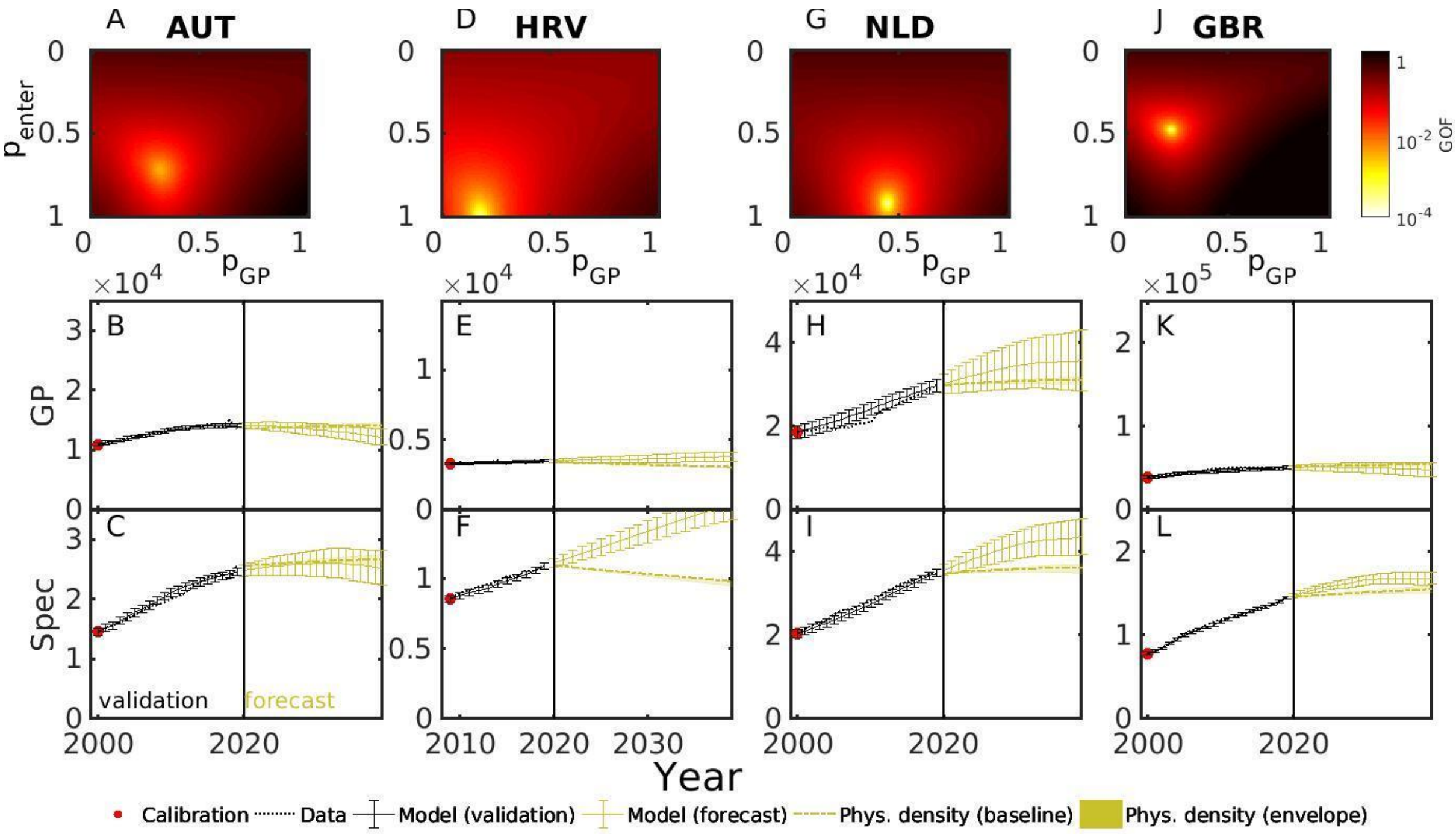

*Figure 2: Output of the minimal model for four different countries. (A) The chi-squared distance $\chi^2$ as goodness-of-fit (GOF) measure between data and model is shown for all settings of $p_{enter}$ and $p_{GP}$ for Austria, indicating a clear global optimum (yellow circle). Results are also shown for the numbers of (B) GPs and (C) Specialists in Austria. The red circle shows the year of calibration, where the stocks of physicians in the model are taken from the data. We then compare the data (black dotted line) with the model (black solid line with error bars) in the validation phase; error bars show the RMSE between data and model In the forecast phase, we show the line of constant physician density in the baseline scenario (yellow dash-dotted line) with a yellow area that envelops all other population growth scenarios. The model timeseries is shown as yellow solid line with errors propagated from the validation phase. The same results are shown for (D-F) Croatia, (G-I) Netherlands and (J-L) Great Britain. At the end of the forecast window, the stock of GPs is substantially lower than the constant density line in (B) Austria and (J) Great Britain; for specialists the projected stocks are above the constant density line for (F) Croatia and (G) the Netherlands. In all other cases, the forecasted supply overlaps with constant density within the margin of errors.*

Whenever the model forecasts lie below the line of constant physician density, the density gap is negative, see for instance GPs in Austria, Figure 2B. If the model forecasts lie above the constant density line, we speak of a positive density gap, as for specialists in Croatia, Figure 2F. The four countries in Figure 2 have been chosen to be representative of what we observe in the other countries for which we show the validation and forecast timeseries of GPs and Specialists in the supplement, Figure S1, Figure S2, and Figure S3. In (B) Austria we see a negative density gap in GPs, like in other countries such as Belgium or France. There is also a negative gap for specialists in Austria (C), though the line of constant physician density falls within the error margin of the forecasted stock of specialists. In (E, F) Croatia we find almost no density gap in GPs but a strongly positive one in specialists that comes from a combination of (i) increasing numbers of specialists in the validation window and (ii) expectations of negative population growth in the future. We find similar results for several Southern and Eastern European countries including Malta, Portugal, Romania, and Slovenia, as well as in the Baltic states with higher margins of error (Estonia, Latvia, Lithuania). The Netherlands (H, I) show a tendency toward positive density gaps in GPs and specialists in combination with positive population growth. Finally, in (K, L) Great Britain the lines of constant physician density fall within the error margins of the forecast for both GPs and specialists (despite negative density gaps), like Denmark, Norway or Germany.

| | ***Density Gaps (RMSE) without Covid shock*** | | | ***Covid shock (Additional physicians per year)*** | | | ***Density Gaps (RMSE) incl. Covid shock*** | | |
|---|---|---|---|---|---|---|---|---|---|
| ***Country*** | ***GP*** | ***Specialist*** | ***All*** | ***GP*** | ***Specia list*** | ***All*** | ***GP*** | ***Specialist*** | ***All*** |
| **Austria** | -6(4)% | -5(9) % | -11(10) % | 18 | 35 | 53 | -8(4)% | -7(9)% | -15(10)% |
| **Belgium** | 0(1)% | 14(7) % | 14(7) % | 14 | 24 | 39 | 0(1)% | 14(7)% | 13(7)% |
| **Bulgaria** | -4(10)% | 38(66) % | 34(67) % | 4 | 21 | 25 | -4(10)% | 36(66)% | 32(67)% |
| **Croatia** | 6(3)% | 42(8) % | 47(8) % | 3 | 9 | 11 | 5(3)% | 40(8)% | 45(8)% |
| **Denmark** | 1(1)% | 11(2) % | 12(2) % | 4 | 8 | 12 | 1(1)% | 10(2)% | 11(2)% |
| **Estonia** | 6(6)% | 24(22) % | 29(23) % | 1 | 3 | 4 | 5(6)% | 22(22)% | 27(23)% |
| **France** | -8(10)% | 5(4) % | -3(10) % | 74 | 93 | 167 | -9(10)% | 4(4)% | -5(10)% |
| **Germany** | -5(18)% | -7(61) % | -12(63) % | 30 | 102 | 132 | -5(18)% | -8(61)% | -13(63)% |
| **Ireland** | -14(10)% | 13(13) % | -0(17) % | 8 | 7 | 15 | -14(10)% | 13(13)% | -1(17)% |
| **Italy** | -0(4)% | 13(10) % | 13(11) % | 21 | 77 | 99 | -1(4)% | 12(10)% | 12(10)% |
| **Latvia** | 8(3)% | 14(13) % | 22(14) % | 2 | 7 | 9 | 8(3)% | 12(14)% | 20(14)% |
| **Lithuania** | 16(13)% | 26(24) % | 42(27) % | 4 | 14 | 19 | 15(13)% | 23(23)% | 39(27)% |
| **Malta** | 6(2)% | 31(4) % | 36(5) % | 0 | 1 | 1 | 5(2)% | 30(4)% | 36(5)% |
| **Netherlands** | 8(12)% | 12(7) % | 20(14) % | 27 | 32 | 59 | 7(12)% | 11(7)% | 18(14)% |
| **Norway** | 0(1)% | 4(5) % | 4(5) % | 2 | 4 | 6 | 0(1)% | 4(5)% | 4(5)% |
| **Portugal** | 20(27)% | 11(27) % | 31(38) % | 34 | 32 | 66 | 18(27)% | 10(27)% | 28(38)% |
| **Romania** | -7(3)% | 34(2) % | 27(4) % | 11 | 34 | 45 | -7(3)% | 33(2)% | 26(4)% |
| **Slovenia** | 8(3)% | 30(9) % | 38(10) % | 1 | 5 | 7 | 7(3)% | 28(9)% | 36(10)% |
| **Spain** | 2(7)% | 7(6) % | 10(9) % | 27 | 76 | 103 | 2(7)% | 6(5)% | 9(9)% |
| **Sweden** | 0(2)% | 5(3) % | 5(4) % | 9 | 31 | 40 | 0(2)% | 3(3)% | 3(4)% |
| **United Kingdom** | -2(2)% | 4(2) % | 2(3) % | 60 | 172 | 233 | -2(2)% | 3(2)% | 0(3)% |

*Table 1: Results for the density gaps with and without the Covid shock for GPs, specialists, and all physicians for each country. Numbers in bracket denote the propagated RMSE. The impact of the Covid shock is also given in absolute numbers of additional physicians per year.*

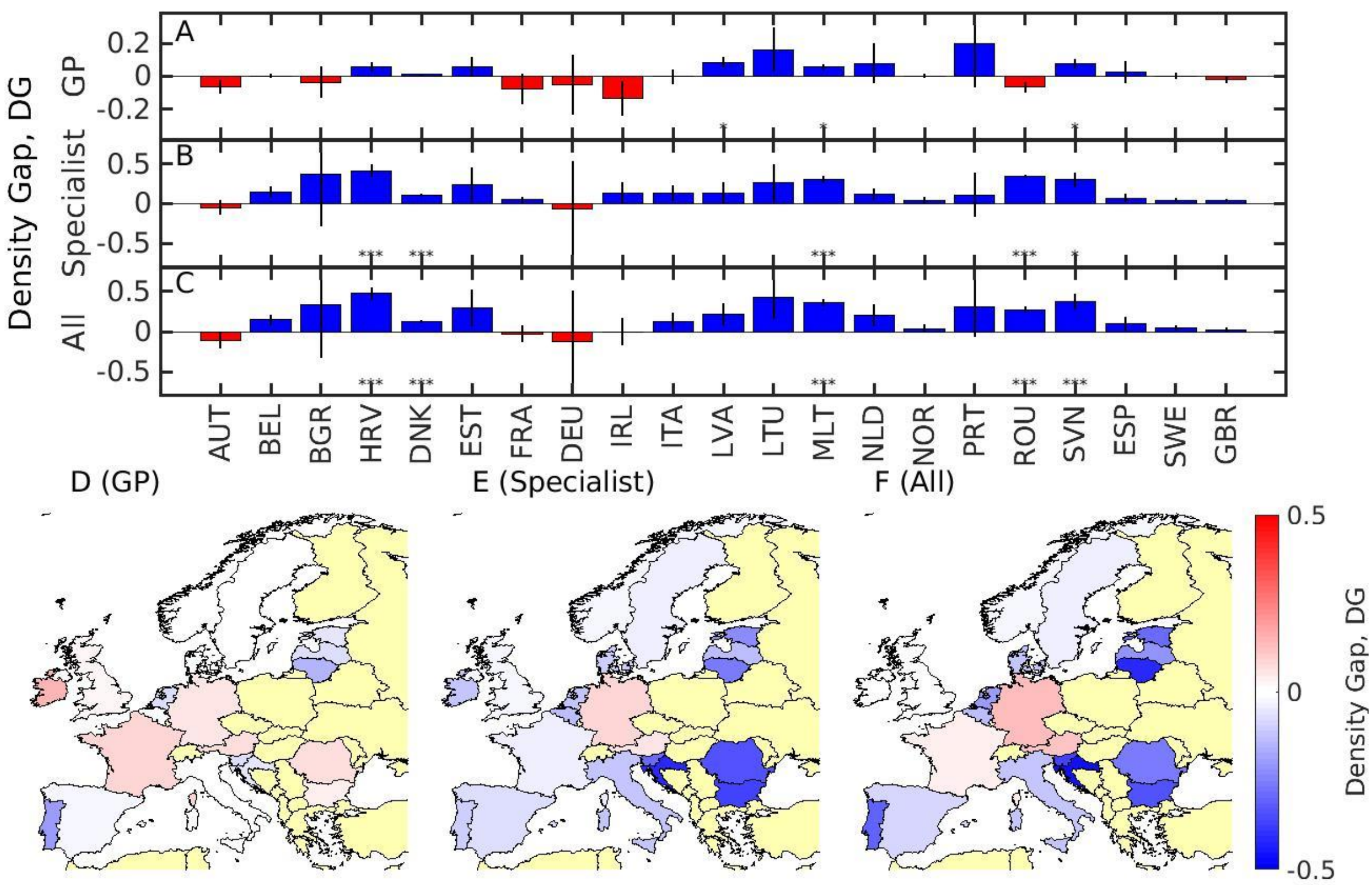

*Figure 3: Results for the density gaps for (A) GPs, (B) specialists and (C) all physicians for each country in the minimal model. Black lines denote the RMSE. Positive (negative) density gaps are shwon in blue (red); gaps that are significantly different from zero are marked by asterisks (* p<0.01, ** p<0.001, *** p<0.0001). A political map with countries colored according to their density gaps is also shown for (D) GPs, (E) specialists and (F) all physicians. Overall, we find negative density gaps for GPs in most countries (except Slovenia and some Baltic states), and positive gaps for all physicians and particularly for specialists that are most pronounced in Southern and Eastern European countries.*

**Density Gaps.** Summary results for the density gaps (measured relative to the number of graduates in each country) are shown in Figure 3 for (A) GPs, (B) specialists and (C) all physicians, along with a p-value for whether the density gap is significantly different from zero or not. We also provide an overview of the density gaps in each country on maps for (D) GPs, (E) specialists and (F) all physicians. The following observations hold. Middle European countries have a tendency toward negative density gap in GPs (though not statistically significant), whereas tendency toward positive density gaps in GPs are mostly found in countries with negative population growth as described above (Latvia, Slovenia), see Figure 3A. For specialists, Figure 3B, we find consistently positive density gaps which are in many countries strongly significant (Croatia, Denmark, Malta, Romania, Slovenia). Considering all physicians, Figure 3C, we still find many significantly positive density gaps in Eastern and Southern European countries (Croatia, Malta, Romania, Slovenia) and in Denmark, whereas in other countries there is no significant overall density gap.

**Extended Model.** Results for the extended model for Austria are shown in Figure S4, where each professionality (GPs and specialists) has been split into three health-care sectors, namely (A,B) contracted, (C,D) employed, and (E,F) non-contracted physicians. Overall, we see similar tendencies as in the minimal model with a stronger negative density gap for GPs compared to specialists. Due to the increased number of graduates in this model version, the density gaps are of smaller absolute value. However, a closer examination shows that the behavior within the individual sectors is quite heterogeneous. For both (A) GP and (B) specialists there is a decline in the number of contracted physicians (driving the negative density gaps) opposed to an increase in non-contracted physicians (E, F), whereas the numbers of employed physicians (C,D) show no significant density gaps.

**Covid Shock.** Table 1 shows how the Covid shock impacts these density gaps for GPs and specialists. Overall, we see modest changes in the density gaps with reductions (less physicians becoming available) ranging from close to zero (e.g., Norway) to maximally 4% (e.g., Austria). These reductions are typically smaller than the uncertainties stemming from the model validation procedure.

# Discussion

In this work we introduced a novel data-driven modeling approach to forecast health workforce supply. The main idea is to formulate a stock-flow-consistent model which requires only net rates of changes in workforce as input such that all model parameters can be estimated directly from historical timeseries. The resulting model has therefore no free parameters and, in its minimal version, can be studied for 21 European countries using public data. We use the model to study a conservative scenario regarding how sequelae and long-lasting symptoms of severe and non-severe Covid might impact the health workforce and the demand for it.

The internal validation of the model can be performed by comparing its output to data. Therefore, we initialize the model such that the numbers of physicians are identical in data and model in the first year only. The model is then forward-iterated until 2019; the last year for which data was available in most countries. Model quality can then be computed as the deviation between data and model, providing us with a natural estimate for how well our model assumptions are compatible with the data. We find that the model fits the data quite well if the data timeseries show consistent trends without severe breaking points. However, for violently fluctuating input timeseries model quality deteriorates, see for instance the results for Estonia, Germany, Ireland, or Lithuania. High (poor) model quality results in low (high) forecast error.

We compared the model forecasts for the physician numbers to the levels that would be required to keep the physician density constant with respect to 2019. The annual differences between the physician stock forecasts and these "isodensity" lines we defined as density gaps. A positive (negative) density gap means that current forecasts point toward an increased (decreased) physician density until 2040. The gaps were measured relative to the number of graduates in the country. The density gaps can therefore be interpreted as the fractions by which the number of annual graduates should be decreased (increased) to keep the physician density constant.

Note that positive or negative density gaps indicate by no means that there is a shortage or oversupply of physicians. Such conclusions could only be drawn under the assumptions that (i) supply equaled demand in 2019 and (ii) future demand will not be impacted by technological, epidemiological or demographic factors other than total population growth. Besides some technical limitations of our approach (see below), one also needs to address the conceptual issue that in certain sectors or countries it might be desirable to increase or decrease the physician density because of changes in demand of health services. We do not address the modeling of demand in the current work, beyond the implemented Covid shock. In our view, the way forward would certainly lie in employing high-throughput machine learning and network analysis techniques to predict prevalence of individual disorders [27, 28] and to relate these trends to the numbers of different types of healthcare provider that patients with a specific disease tend to utilize [29]. This clearly goes beyond the scope of this article.

For GPs we find mixed patterns of results for the density gaps. There are countries with tendencies toward negative (Austria, France, Ireland) and positive (Lithuania, Portugal) gaps. In contrast, all density gaps significantly different from zero for specialists are positive. Overall, the positive gaps for specialists outweigh the negative ones for GPs, meaning that all significant results for the density gaps for all physicians are again positive. Baltic, Southern and Eastern European countries with a projected negative population growth are particularly likely to show such positive gaps (Croatia, Malta, Romania, Slovenia). This might be surprising at first glance, as most of these countries are usually associated with physician shortages and substantial emigration of physicians [30, 31]. However, the main driver of these positive density gaps is the negative population growth combined with persistent increases in medical specialists. Indeed, many of these countries show a skew towards higher proportions of specialist physicians with 24% in Slovenia [33] (compared to around 30% as the European average). The health workforce for Slovenia has been projected to grow to a number around 8,000 until 2040 (no margin of error given) [34], whereas we estimate a growth to around 9500(900).

In terms of external validation of the model, we can compare our findings to results from other countries that do employ sophisticated quantitative models to forecast supply in their health workforce. For instance, in Belgium there is a known specialty imbalance between general and specialist physicians [35] which is reproduced in our model too. The Netherlands experienced a shortage of physicians around 2000 which was followed with a health-care reform and the financial crisis that resulted in a slower increase in healthcare demand [21]. There have now even been reports of increasing unemployment amongst specialists [36], hinting at a so-called “pork cycle” in the labour market between over- and undersupply [4]. We find also positive density gaps for general and specialist physicians in the Netherlands. Overall, in most of the cases our findings suggest that the trend of increasing supply of physicians is expected to continue over the next decades. These increases are expected to be stronger for specialists than for general physicians. Between 2030-2040 the growth in supply levels off in many countries including Austria, Denmark, Netherlands, or Great Britain.

To summarize these findings, we see that most countries should be able to replace exiting (retiring) physicians up to a point where their overall physician density will not significantly decrease. However, we find consistent signs for a suboptimal distribution between general and specialist physicians, where positive density gaps for specialists typically outweigh negative ones for GPs. As already mentioned above, positive density gaps do not indicate the absence of a potential physician shortage or even a surplus. For such a diagnosis one would also need to take other factors like changes in demand properly into account. These factors include epidemiological changes such as the epidemics of chronic disorders that is compounded by population ageing [10]. Furthermore, to determine a potential over- or undersupply, it is not enough to only count heads. The health workforce is getting younger and more female which could potentially impact the number of hours that physicians will work on average [16]. All these factors typically mean an increased supply of physicians will be necessary to meet the population's increasing demands.

We find a modest impact of the Covid shock on these density gaps. Across different countries, the Covid shock decreased the density gaps in a range from 0% to 4%. Central to this result is the assumption that individuals with Covid have a higher risk of requiring outpatient care with a hazard ratio of 1.2 (95% confidence interval 1.19-1.21) [12]. The differences in reductions of physicians between the countries only depend on the case numbers as well as (if available) weekly hospital and ICU admissions. We assume the same rate of detection as well as survival, and hospital and ICU admission rates (if the data is not available). Furthermore, we make the conservative assumption that European countries will experience similar surges in severe and non-severe Covid cases in the upcoming years as experienced in 2020 and 2021. If these outcomes were to be attenuated by a given percentage over the coming years, the measured changes in density gaps and additional physicians would decrease proportionally. However, even under the worst-case scenario of no attenuation the changes in density gaps due to Covid are typically smaller than the uncertainties associated with their estimation. Given the currently available evidence regarding the prevalence of long-lasting Covid sequelae and symptoms, we conclude that it cannot be expected to lead to substantial impacts on overall demand for physicians.

For the specific case of Austria, we showed how using additional input data the minimal model can be extended to a multi-sectoral and multi-professional setting. Compared to other EU countries, Austria has the second largest hospital-bed-per-population (after Germany) and the second highest physician density (after Greece) [37]. Patients in Austria can freely choose to access contracted or non-contracted physicians, where in the latter case they are reimbursed for 80% of what the insurance would have paid for contracted care. In the extended model, we distributed the generalist and specialist physicians to these three main sectors, namely contracted, employed (hospital) and non-contracted physicians. In addition, we consider a scenario with dynamic parameter settings in the extended model. First, we include projections for additional graduates because of a new medical faculty. Second, we include the trends of how the proportions in the individual sectors and types of physicians are changing over time. As a result, the overall density gap closes from about -16% in the minimal to -7% in the extended model.

We found strongly varying results across the individual sectors. For employed physicians, the line of constant physician density falls within the margin of error of the projected numbers for GPs and specialists. For non-contracted physicians we found positive density gaps of smaller absolute value than the negative gaps for contracted doctors. The results therefore indicate not only an imbalance between generalist and specialist physicians, but also between contracted and non-contracted physicians.

The strategy outlined above to generalize the single- to a multi-sector model can also be employed to extend the approach to cover multiple regions and/or additional types of providers. In the current work we exclusively use publicly available data. However, in most countries data on additional stocks of healthcare providers might be available and could therefore be used to refine the model results. A "wish list" of data availability for each type of provider to include in the model would consist of (i) information on age and sex distribution of the current personnel and (ii) age- and sex-specific retirement rates (though the latter could also be estimated from longitudinal information on the age and sex distribution, as we did here). Of importance would be to extend the current approach to a multi-regional setting in order to address potential geographic imbalances between urban and rural areas, which are assumed to be an important issue in several European countries [4, 35].

In summary, we have presented a novel approach to forecasting supply in health workforce. We presented a parameter-free model, fully calibrated from publicly available data for 21 European countries, that can be extended to multi-regional, multi-sectoral and multi-professional settings. We circumvented the curse of high dimensionality that plagues many other approaches by focusing on net rates of changes that can be estimated from data. However, we ensured that those neuralgic points that are used by policy makers to steer supply remain explicit in the model. This includes (i) the number of graduates, a "knob" that can be turned for instance by a *numerus clausus*, and (ii) the preferences of these graduates to enter specific fields. While many countries so far focused mainly on regulating their numbers of graduates, our results call further attention to the issue of combatting imbalances of physicians across sectors and medical fields. Further, based on currently available evidence we conclude that long-lasting sequelae and symptoms of Covid are not likely to lead to unexpected surges in physician demand.

## Methods

**Data**. The minimal model is exclusively based on publicly available data from EUROSTAT[1]. We use the number of physicians by age and sex (table hlth_rs_phys) and by medical specialty (hlth_rs_spec), the number of graduates (hlth_rs_grad) and information on health workforce migration (hlth_rs_wkmg). We consider the timespan 2000-2016 and only include those European countries where data on the number of physicians by specialty has been available from at least 2009 onwards. Population forecast scenarios are taken from the EUROSTAT table "proj_15npms". For the extended model in Austria, we additionally use information on the number of physicians per sector. The numbers of contracted and non-contracted physicians for the years 2011 and 2015 were obtained from answers of parliamentary questions[2], the numbers of employed physicians by specialty are available from Statistik Austria[3].

---

[1] https://ec.europa.eu/eurostat/data/database, accessed 07/19/2019.

**Age distributions and exit rates**. From the data we obtain $X_i^{(c)}(s,g,t)$ as the number of physicians of sex $s$ (male/female) and age group $g$ in year $t$ that are working in field $i$ in country $c$. In the minimal model we consider two fields (GPs, specialists), in the extended model six (contracted, employed, non-contracted GPs and specialists, respectively). In the following, we will suppress the country index. Age information of the physicians is given in ten-year groups. We estimate the numbers for age years $a$, $X_i(s,a,t)$, by assuming a linear change in numbers of physicians per age year between two adjacent age groups, see the black lines in Figure 1 (C, D). We next compute the net rates of change, $\alpha_i(s,a,t)$, as $\alpha_i(s,a,t) = \big(X_i(s,a+1,t+1) - X_i(s,a,t)\big)/X_i(s,a,t)$. From this we obtain the effective exit rate, $\gamma_i$, as $\gamma_i(s,a,t) = \alpha_i(s,a,t)$ if $\alpha_i(s,a,t) < 0$ and $\gamma_i(s,a,t) = 0$ otherwise. In the data, we observe these exit rates for each sex, age, and year, but unfortunately not for the individual fields (hlth_rs_phys only gives age and sex data for all physicians). In the model we therefore assume that the exit rates only depend on age and sex but not on field. This means we use the same average value of $\gamma_i(s,a)$ for each field $i$, where the average has been taken over all years $t$ for which there is data in a country, see also Figure 1 (E, F). We assume that $\gamma_i(s,a)$ remains constant during the forecast window (2020-2040). For countries which do not report age- and sex-specific stocks of physicians, we use reference exit rates and reference sex and age distributions of the stocks instead. These reference values are computed as described above from the aggregated stocks of all countries with complete data.

**Model in-flow, graduates and migration**. New physicians enter the health-care system via one of two processes. They might enter through graduation or migration, see also Figure 1 (A, B). In both cases we assume that they are young enough when they enter that they do not exit (retire) before the end of the forecast window (2040). For the forecast window, we assume an in-flow equal to the average in-flow observed in the years 2014-2019, if not stated otherwise. The sex distribution of graduating or immigrating physicians is assumed to follow the same distribution as observed in the age group 25-34y in 2019 and assume that the new physicians enter in this age group. We denote the number of graduates and migrants of sex $s$ and age $a$ in year $t$ by $Y(s,a,t)$ with $Y(t) = \sum_{s,a} Y(s,a,t)$.

**Model initialization**. As calibration year for a country, we use the first year, $t_0$, in the observation window where data is available and no break in the data has occurred (according to the data source). In most countries, this is the year 2000. Let $N_i(s,a,t)$ be the number of physicians in field $i$ of sex $s$ and age $a$ in year $t$. The model is initialized by setting $N_i(s,a,t_0) = X_i(s,a,t_0)$.

---

[2] Parliamentary question No. 10106/J to the Austrian Parliament.
[3] https://www.statistik.at/web_en/statistics/index.html, accessed 07/19/2017.

**Model parameters**. The following model parameter need to be specified in the model. First is the probability that a graduating student enters as a physician, $p_{\text{enter}}$. Second, for each field $i$ there is a probability $p_i$ that it will be chosen by a new physician. In the minimal model, physicians choose to become GPs with probability $p_{\text{GP}}$ and specialists with $1 - p_{\text{GP}}$.

**Model protocol**. The model dynamics is completely specified by the following protocol that advances the model from year *t* to year *t*+1.

1. Physicians age; set $N_i(s, a + 1, t + 1) = N_i(s, a, t)$.
2. Physicians with sex *s* and age *a* exit with probability $\gamma_i(s, a)$ in field *i*.
3. New physicians enter, $N_i(s, a + 1, t + 1) = p_{\text{enter}} p_i Y(s, a + 1, t + 1)$.

The model dynamics is therefore given by the update equation $N_i(s, a + 1, t + 1) = (1 - \gamma_i(s, a))N_i(s, a, t) + p_{\text{enter}} p_i Y(s, a + 1, t + 1)$.

**Model calibration using goodness-of-fit (GOF)**. The parameters $p_{\text{enter}}$ and $p_i$ are estimated by comparing the model output with historical data from the observation window between $t_0$ and 2016. Let us denote the total number of physicians in field *i* in 2016 in the data by $Z_i = \sum_{a,s} X_i(s, a, 2016)$. This is compared to the model result $M_i = \sum_{a,s} N_i(s, a, 2016)$ The weighted sum of the chi-squared distance between data and model timeseries is computed as follows. Let $w_i$ be the proportion of physicians that are active in field *i* in 2016, $w_i = Z_i / \sum_i Z_i$. The weighted sum of the chi-squared distance over all fields is then $\chi^2 = w_i((Z_i - M_i)/Z_i)^2$. To calibrate the minimal model, we perform a brute-force search over all possible values of $p_{\text{enter}}$ and $p_{GP}$ (in increments of 0.01 between the lower and upper bounds of zero and one) to find the parameter setting that leads to the smallest value of $\chi^2$. In the extended model, see also SI Note S1, the brute-force strategy is not advisable anymore because of the increased dimensionality. Due to the smoothness of the objective function, an optimum can be better identified by means of stochastic gradient descent or related methods.

**Density gaps**. The model output for a given field is compared to the number of physicians that would be required to keep the number of physicians per population constant at levels of 2016, using the baseline population growth scenario provided by EUROSTAT for *T* years in the future (forecast window of *T* years). Assume that $C_i(t_0 + T)$ physicians would be required after *T* years for a constant density. The density gap *DG* is the difference in number of physicians between $C_i(T)$ and $M_i(t_0 + T)$ per year and physician inflow, $DG(T) = (M_i(t_0 + T) - C_i(T))/TY(t_0 + T)$.

**Covid Shock.** The decrease in physician supply and increase in population demand due to the Covid pandemic is estimated based on several assumptions. First, we assume that all surviving, severe cases (Covid patients who were treated in ICU), increase the demand and that a certain fraction of non-hospitalized Covid patients additionally increases the demand due to long-term consequences of a Covid infection. The number of total cases for each country as well as, if available, information on hospital and ICU admissions per week, is obtained from publicly available databases [38]. If the rate of severe cases is not available for a country, we use the weekly hospital admissions and assume that 37% go into ICU [39]. If neither one of these datasets is available, we assume that 10% of the total cases are admitted to the hospital [40]. We set the rate of ICU survival to 70% [39]. In order to calculate the increase in demand due to long-term consequences of a Covid infection, we assume that for each detected there is also an undetected case, which applies only to non-hospitalized cases. The burden of long-term consequences on the physicians is implemented by calculating the increase in outpatient encounters. The increased risk to require outpatient care in case of an infection is 20% (hazard ratio of 1.2 with 95% confidence interval 1.19-1.21) [12]. Based on these assumptions, we estimate the demand increase for each country due to long-lasting Covid symptoms and sequelae.

For the estimation of decrease in physician supply, we take the rate of severe cases per country and assume that 11.4% are unable to work [14]. The rates for increase in demand and decrease in supply are implemented in the model in the projections of population and physicians, respectively. We calculate the additional physicians needed compared to the scenario without Covid and the density gaps for every country.

## Acknowledgments

We acknowledge support from the WWTF "Mathematics and…" project MA16-045 and FFG project 857136. We thank Ines Czasny, Michael Hummer, and Gunter Maier for helpful discussions.

# Supplement

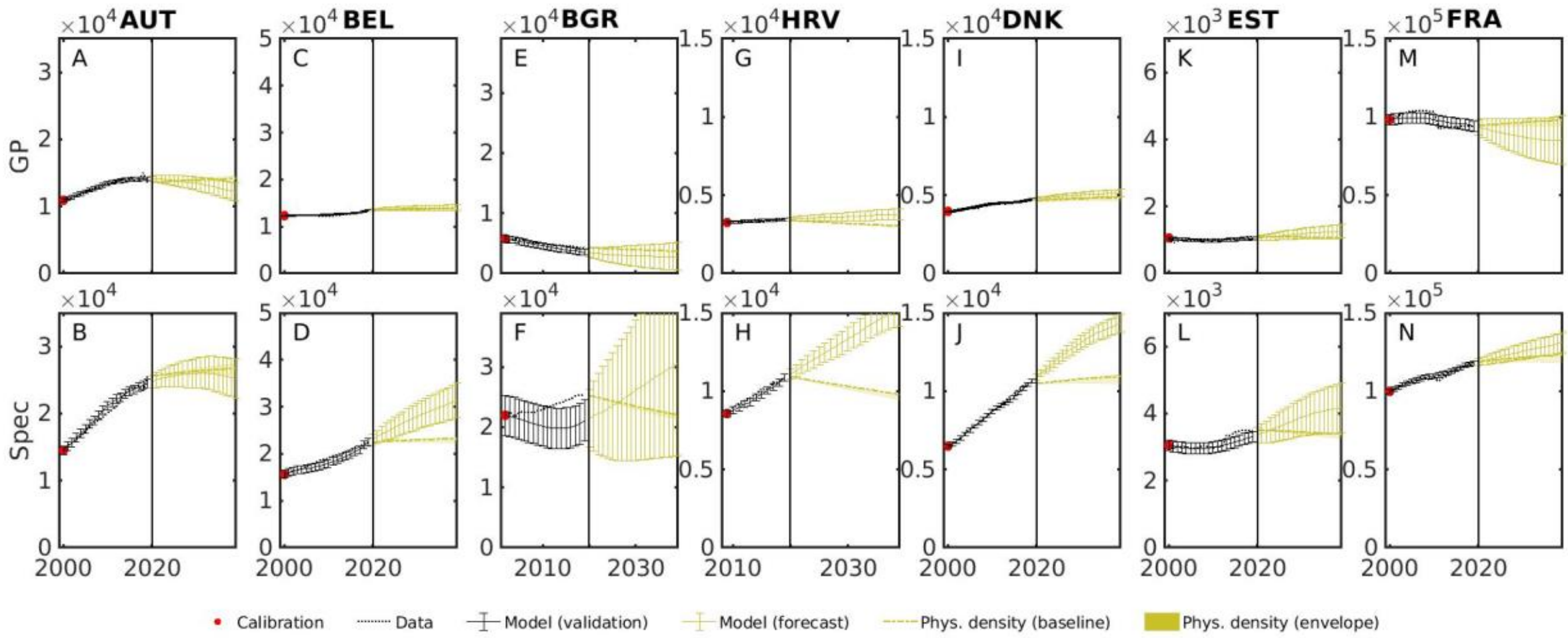

*Figure S1: Model results for (A, B) Austria, (C, D) Belgium, (E, F) Bulgaria, (G, H) Croatia, (I, J) Denmark,(K, L) Estonia, and (M, N) France. See Figure 2 for a description of the visual coding.*

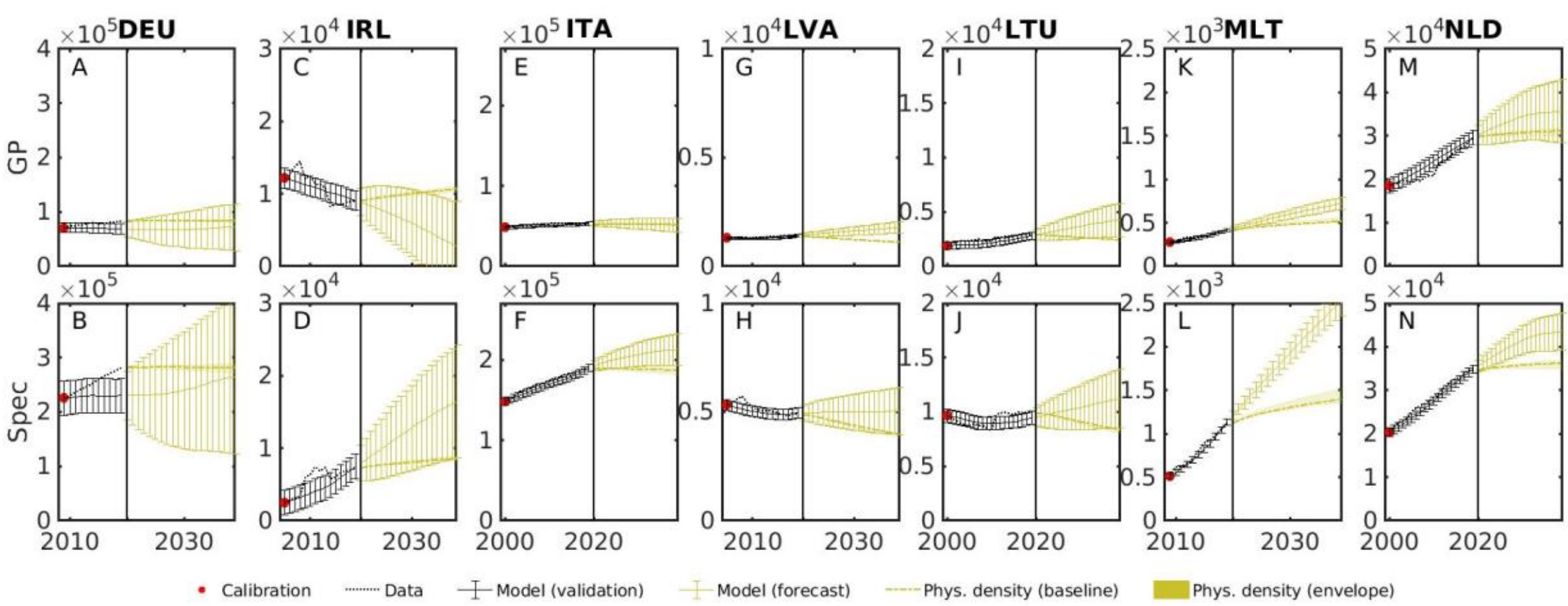

*Figure S2: Model results for (A, B) Germany, (C, D) Ireland, (E, F) Italy, (G, H) Latvia, (I, J) Lithuania,(K, L) Malta, and (M, N) Netherlands. See Figure 2 for a description of the visual coding.*

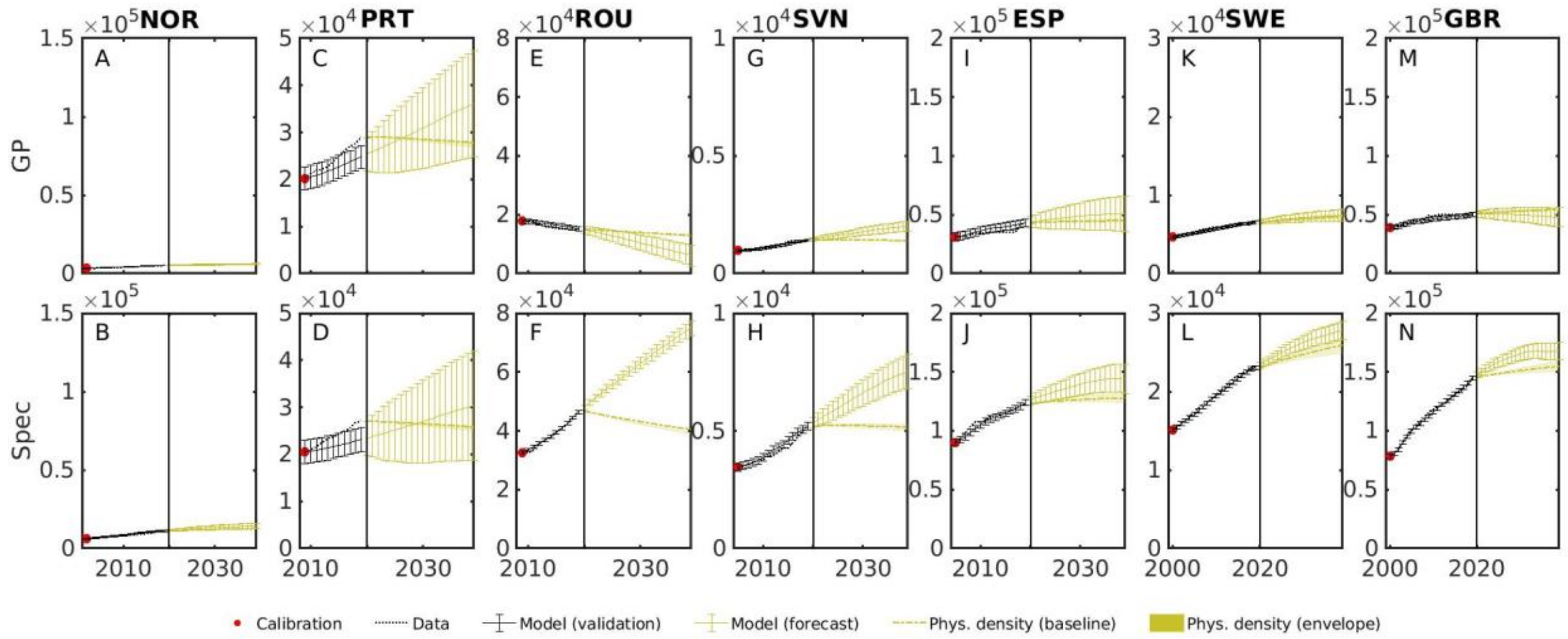

*Figure S3: Model results for (A, B) Norway, (C, D) Portugal, (E, F) Romania, (G, H) Slovenia, (I, J) Spain,(K, L) Sweden, and (M, N) Great Britain. See Figure 2 for a description of the visual coding.*

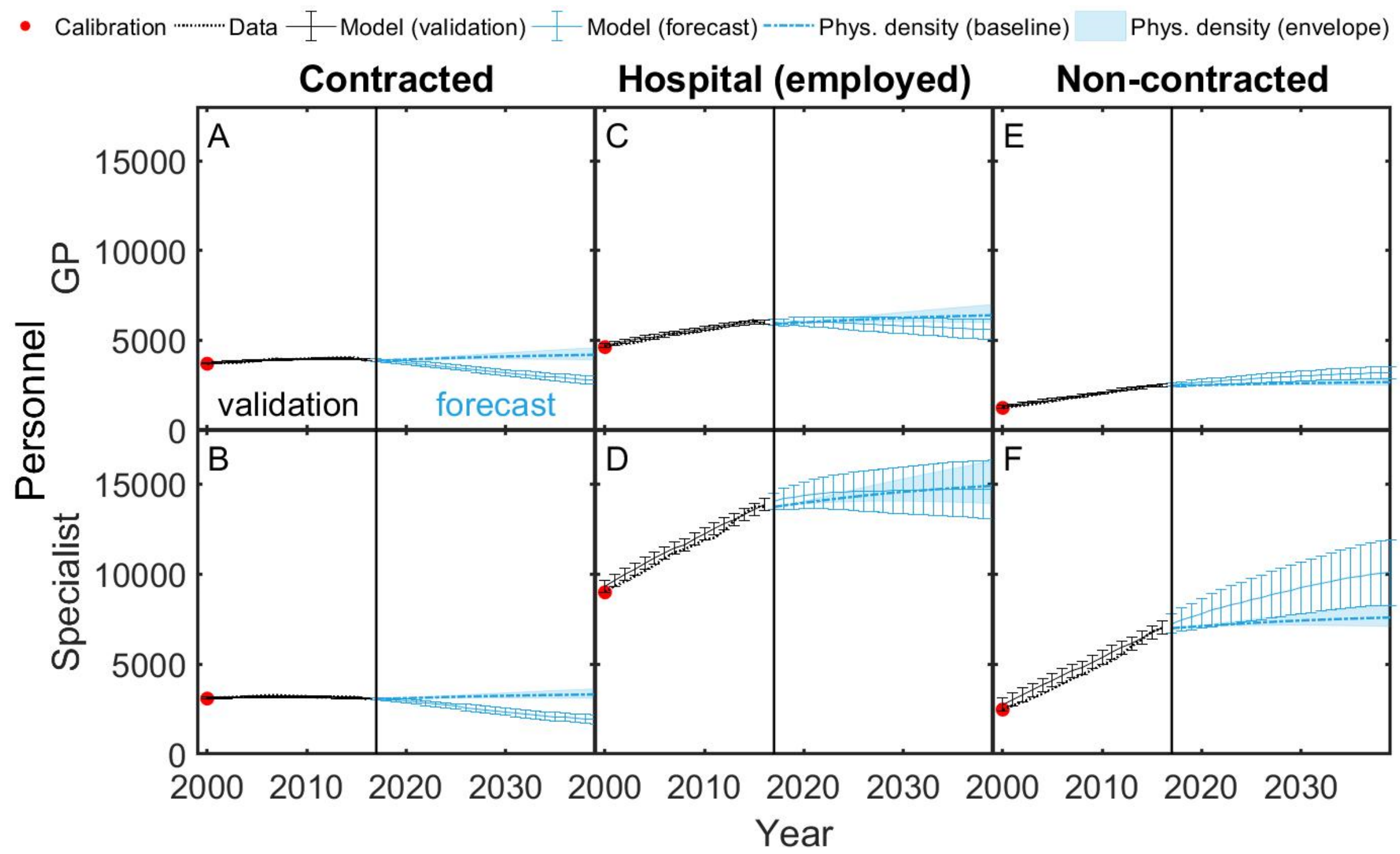

*Figure S4: Results of the extended model in Austria. We use the same visual coding as in Figure 2. There is a negative density gap for contracted (A) GPs and (B) specialists, no substantial gap for employed (C) GPs and (D) specialists, and a positive density gap for (E,F) non-contracted physicians.*

### SI Note 1: Description of the extended model

Given more detailed data on health personnel, the minimal model can be extended to cover multiple health sectors and regions. We show this with the example of Austria where physicians (GPs and specialists) belong to one of three sectors. They are either counted as contracted (they have a contract to bill their services directly with social security institutions), employed (most of which work in a hospital) and non-contracted physicians (have no contract with social security institutions and patients have to pay privately for consultations, though some of these costs might be eligible for reimbursement by the insurances). Data for the split of Austrian GPs and specialists into these three sectors is available for 2012 and 2016. To account for changes in the proportion of doctors in each sector and professionality, we computed the linear trend for each sector for GPs and physicians and extrapolated this trend for the time window 2000-2040. Instead of two different types of physicians as in the minimal model, we now have six different types. The variable $p_{\mathrm{GP}}$ is replaced by a 3-by-2 matrix $P$ (rows correspond to sectors, columns to GPs and specialists), the entries in $P$ sum to one. The elements in $P$ are now cumbersome to be estimated by brute force, so we employ a heuristic gradient method, see Methods.

With the extended model we also show how certain interventions can be formulated within the model. Since 2014 Austria has a new medical faculty from which the first graduates are expected around 2020 when they start there “Turnus”. It is expected that the number of graduates per year will reach its full capacity of 300 until 2029. In the extended model, we can study the impact of this new faculty by adding these graduates (assuming a linear growth form zero in 2019 to 300 in 2029) to the input timeseries shown in Figure 1A.